# Move of a large but delicate apparatus on a trailer with air-ride suspension

B. Thomas[1], D. Will[1], J. Heilman[1], K. Tracy[1], M. Hotz[1], D. Lyapustin[1], L. J Rosenberg[1], G. Rybka[1], A. Wagner[1], J. Hoskins[2], C. Martin[2], N. S. Sullivan[2], D. B. Tanner[2], S. J. Asztalos[3], G. Carosi[3], C. Hagmann[3], D. Kinion[3], K. van Bibber[4], R. Bradley[5], J. Clarke[6]

[1]University of Washington, Seattle Washington 98195
[2]University of Florida, Gainesville Florida 32611
[3]Lawrence Livermore National Laboratory, Livermore California 94550
[4]Lawrence Livermore National Laboratory/Naval Postgraduate School, Monterey California 93943
[5]National Radio Astronomy Observatory, Charlottesville Virginia 22903
[6]University of California and Lawrence Berkeley National Laboratory, Berkeley California 94720

*When valuable delicate goods are shipped by truck, attention must be paid to vibrations that may cause damage. We present a case study of moving an extremely delicate 6230-kg superconducting magnet, immersed in liquid nitrogen, from Livermore, CA to Seattle, WA showing the steps of fatigue analysis of the load, a test move, and acceleration monitoring of the final move to ensure a successful damage-free transport.*

## I. Introduction

Vibration during truck transport may cause damage to delicate goods. Before transporting a unique valuable object, preliminary analysis of the object and vibration tests under the move conditions can provide confidence that the object will not be damaged during shipping. A case study of moving an extremely delicate 6230-kg superconducting magnet along the west coast of the United States follows.

In the summer of 2010, the Axion Dark Matter Experiment (ADMX) was relocated from Laurence Livermore National Labs (LLNL) in Livermore, California to the Center for Experimental Nuclear Physics and Astrophysics (CENPA) at the University of Washington, Seattle. The experimental apparatus consists of a large cryostat housing a 6230-kg superconducting magnet. The mass of the entire apparatus is approximately 8000 kg. The magnet normally operates at 4.2 K, the boiling point of liquid helium, but was immersed in liquid nitrogen, at 77 K, during the move. The purpose of shipping at this low temperature was to minimize thermal expansion of the interior supporting structure. The total driving distance was 1400 km.

The cryostat is a stainless steel cylinder approximately 4 m tall and 1.6 m in diameter, providing support and insulation for the superconducting magnet [1]. The magnet is suspended inside the cryostat by three narrow steel rods and is restrained from horizontal movement by 4 fiberglass bands attached to the bottom of the magnet (See Fig. 1(b)). A failure of these supports or restraints could result in damage to the main coil, necessitating costly repairs to the magnet.

To avoid damaging the load, we analyzed primary failure modes and measured the power spectral density (PSD) of the acceleration [2] during a 200 km test run of a flat bed trailer with air ride suspension and a dummy load (Fig. 1(c)). During the actual move we monitored the acceleration PSD of the cryostat to confirm accelerations were within the desired range.

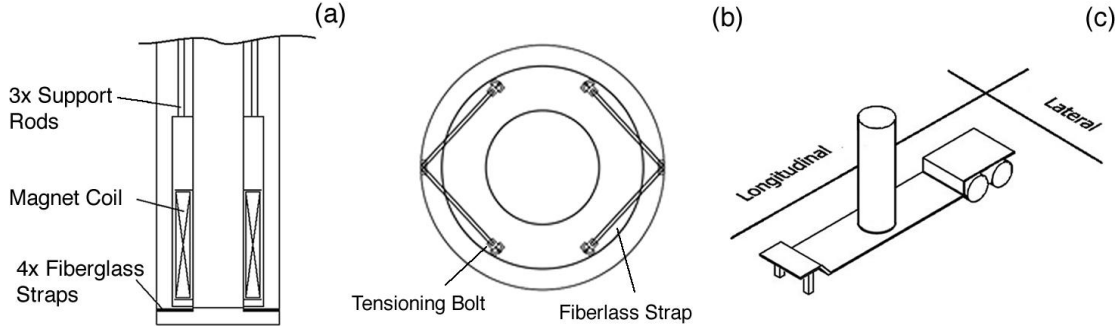

Fig 1: (a) Diagram of ADMX cryostat. (b) View looking up the cryostat bore, showing the fiberglass straps and tensioning bolts securing the magnet. (c) Diagram showing the orientation of the cryostat and trailer axes.

## II. Load Analysis

The primary mechanical resonance of the cryostat was from pendulum motion of the magnet, with the bulk of the restoring force coming from the four fiberglass restraining bands. We model the magnet as a simple harmonic oscillator with mass $m = 6230$ kg and a spring constant $k = 1.7 \times 10^7$ N/m [3] provided by the restraining bands. The resonant frequency was calculated to be

$$\nu_0 = \frac{1}{2\pi} \sqrt{\frac{k}{m}} \approx 10\,Hz.$$

Excessive horizontal accelerations at this frequency could have caused catastrophic failure. This mode was insensitive to vertical accelerations.

We were most concerned with the failure of the stainless steel bolts that tension the fiberglass straps. This failure could occur for two reasons: a yield failure resulting from a large acceleration that surpassed the tensile strength of the bolts, or a large number of smaller accelerations at the resonant frequency, causing the bolts to work harden and fail from fatigue.

### Yield Failure

The tensioning bolts are ½" diameter 316L stainless steel. This alloy has a tensile strength of about 1.2 GPa [4] near the temperature of liquid nitrogen giving a maximum force on the tensioning bolt of 152,400 N.

### Fatigue Failure

Given the nature of road transport, fatigue failure was our greatest concern. Based on the resonant frequency and an assumption that the trip would require about 40 hours of driving time, we estimated that the magnet would go through approximately 1.5 million cycles. Under these conditions, we estimate the fatigue strength of 316 L stainless steel is approximately 35% of its tensile strength [5] giving a maximum force on the tensioning bolt of $F_{max} = 53,340$ N for fatigue failures.

The force on the tensioning bolt depends on the resonant frequency of the magnet, $\nu_0$, the Q factor of the cryostat, and the frequency with which the outer cryostat shell is accelerated, $\nu$. The primary damping was due to the air-ride suspension of the trailer with a typical Q factor of 2 [6]. The PSD threshold for the acceleration of the cryostat shell as a function of angular frequency is given by [7]

$$PSD_{Threshold} = |F_{max}|^2 \left( \frac{(\omega_0^2 - \omega^2)^2 + \left(\frac{\omega_0 \omega}{Q}\right)^2}{k^2 + \left(\frac{m\omega_0\omega}{Q}\right)^2} \right). \quad (1)$$

This threshold represents the maximal magnitude of acceleration that the cryostat can withstand without exceeding the fatigue failure threshold of the tensioning bolts.

Equation 1 shows that the cryostat is most sensitive to accelerations around the resonant frequency of 10 Hz where the fatigue failure threshold drops to $2.6 \times 10^{-3}$ $g^2$/Hz, where g is the acceleration due to gravity. The response of the cryostat quickly becomes less sensitive to accelerations at higher frequencies.

## III. Test Move

The acceleration PSDs of tractor-trailers can vary significantly [6]. To ensure the acceleration threshold would not be exceeded, a test move was designed to measure the accelerations applied directly to the cryostat shell during the transport. A 12-ton forklift (Fig. 2) was chosen as a test mass because its center of mass is similar in height to that of the cryostat. The forklift was loaded onto a Double-Drop Air-Ride Low-Boy trailer, and instrumented with an array of accelerometers near its center of mass. The accelerometers measured the vertical, longitudinal and lateral accelerations at a sampling rate of 512 Hz; the data were stored in Omega TSR-101 and OM-CP-SHOCK101 data loggers for later analysis. Accelerations were recorded for a single day of driving under various road conditions.

Figures 3 and 4 show data from a typical section of what was found to be the most promising driving style. This was achieved by allowing the driver of the pilot car to communicate his assessment of upcoming road conditions with the drivers of the tractor-trailer. Jointly, the drivers dynamically altered their driving speeds based on their experience and expertise. This method yielded the lowest number of dangerous accelerations and minimized the likelihood of damaging the cryostat. We found that a driving speed of approximately 40 mph was optimal for the move. Speeds much greater than this did not produce a significant increase in the acceleration PSD; however, there was insufficient time for the driver of the trailer to react to warnings of changing road conditions coming from the driver of the lead car. Speeds much less than 40 mph showed no significant reduction in the acceleration PSD.

Figure 4 shows the acceleration PSD for the lateral, longitudinal, and vertical axis, and the threshold failure curve. While a number of high instantaneous acceleration events were recorded, accelerations in the range of the resonant frequency of the cryostat from these transient events were well within the tolerance of the yield failure mode. Fatigue failure, which was a larger concern, was within threshold by a safety factor of ten. Table 1 summarizes the maximum and average accelerations along the lateral, longitudinal and vertical axes.

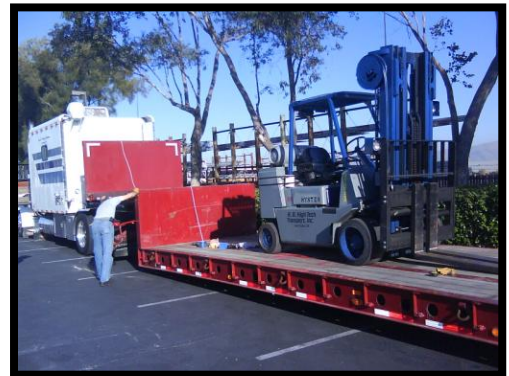

Fig 2: A 12-ton forklift used as a substitute mass for the test move

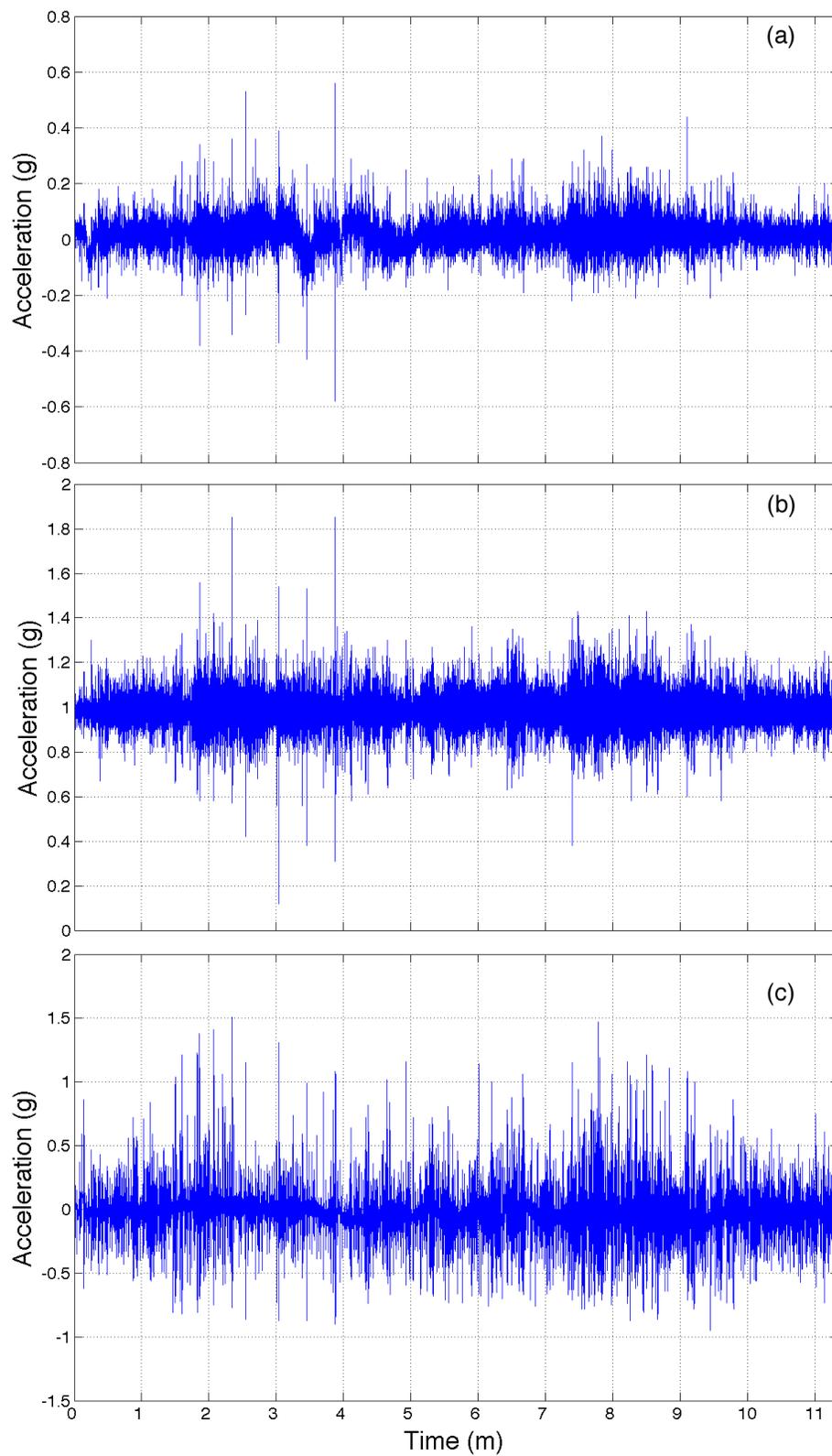

Fig. 3: Time series of (a) lateral (b) vertical and (c) longitudinal axis acceleration for a forklift driven on a tractor-trailer at 40 mph under typical road conditions near Oakland CA.

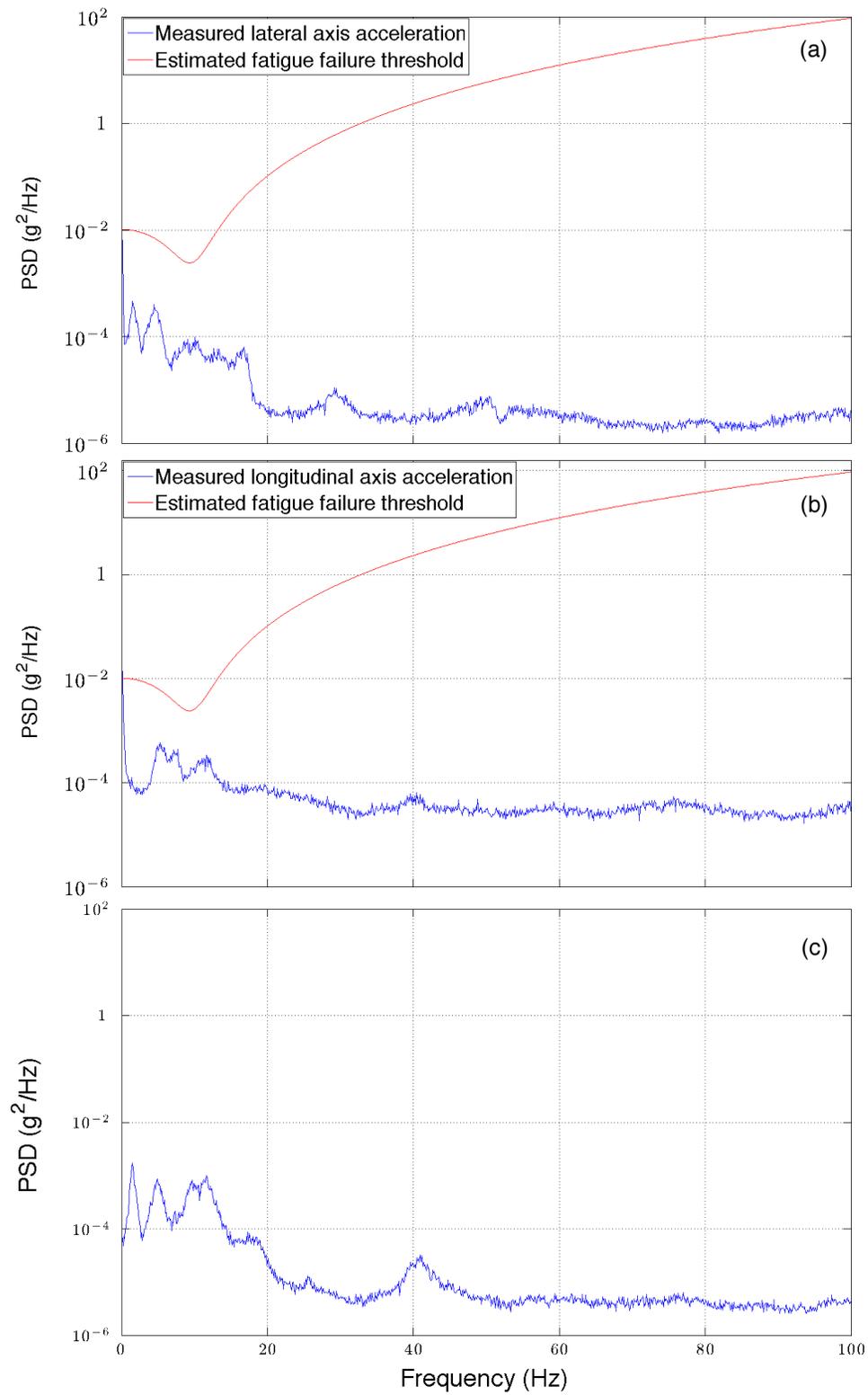

Fig. 4: Acceleration PSD of (a) lateral, (b) longitudinal and (c) vertical axis acceleration for typical road conditions during test move driven at 40 mph near Oakland CA. Figures (a) and (b) also show the PSD form of the acceleration at the fatigue failure threshold.

## IV. The Move

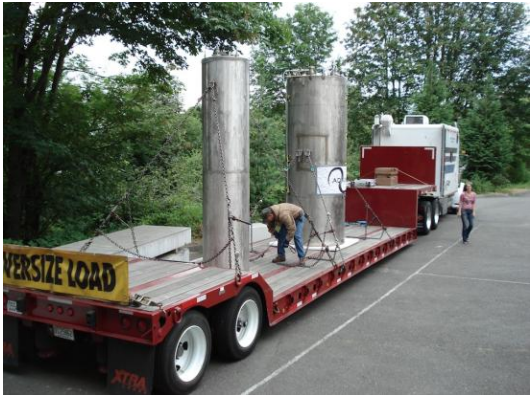

Fig 5: The cryostat and trailer after arriving at the University of Washington. The smaller diameter cylinder in the foreground housed a non-critical experimental apparatus.

The cryostat was moved using the same type of trailer as in the test move, and instrumented in a similar fashion (Fig. 5). A combination of high frequency TSR-101-Transient, and lower frequency OM-CP-Ultrashock continuous Omega data loggers were used to acquire and store the acceleration data. A support crew inside the cabin of the trailer monitored a laptop that displayed live acceleration values and kept a running log of notable events. As in the test move, a lead car was used to notify the crew in the trailer of upcoming road conditions so that speed and driving style could be adjusted.

Figure 6 shows annotated time series plots of the lateral and longitudinal axes. Notable events such as potholes, road conditions and times when the trailer was stopped are labeled as observed by the support crew.

Figures 7 and 8 show PSDs of both rough and smooth road conditions from a typical section of data from the move. As in the test move, accelerations in the range of the resonant frequency are well within the tolerances of the yield and fatigue failure modes. The large maximum accelerations in the lateral and longitudinal axes shown in Table 2 are too short-lived to pose a threat to the cryostat.

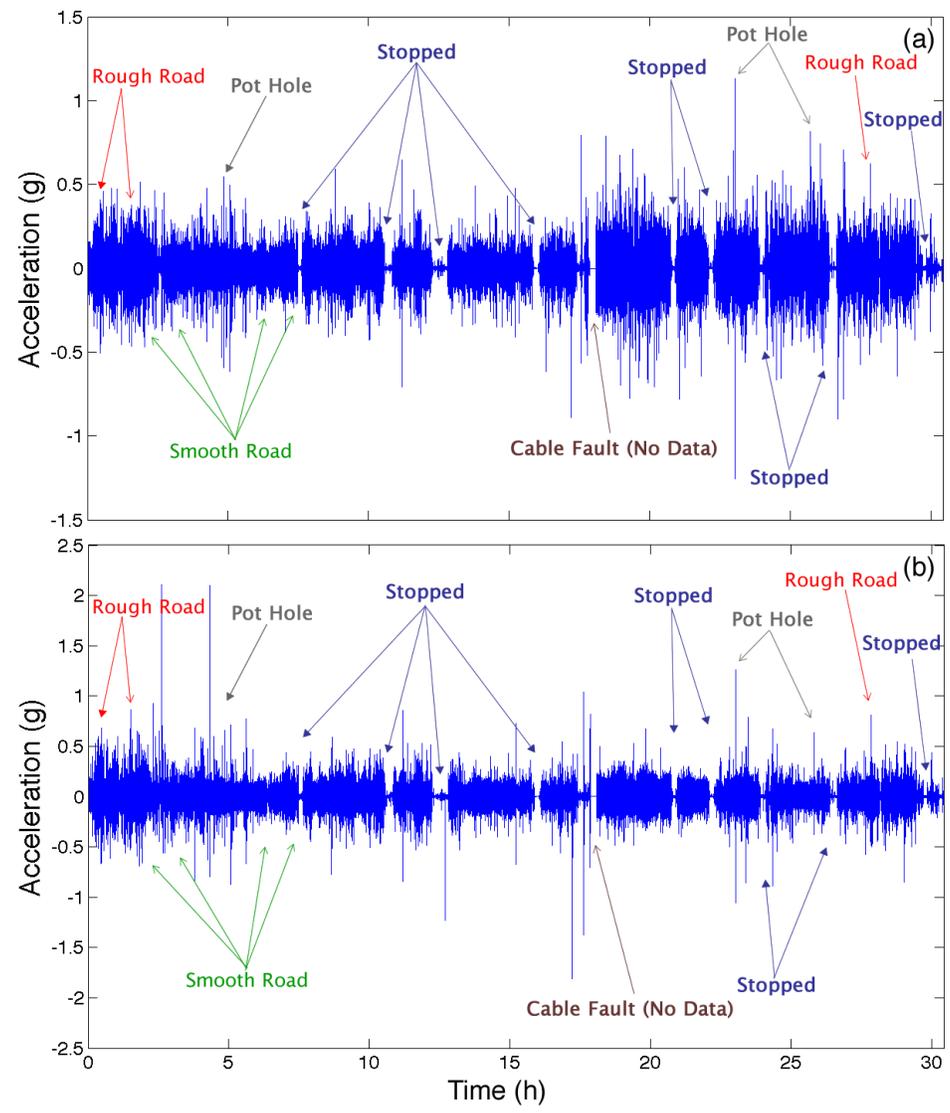

Fig. 6: Annotated time series plot of (a) lateral and (b) longitudinal acceleration over the move from Livermore CA to Seattle WA.

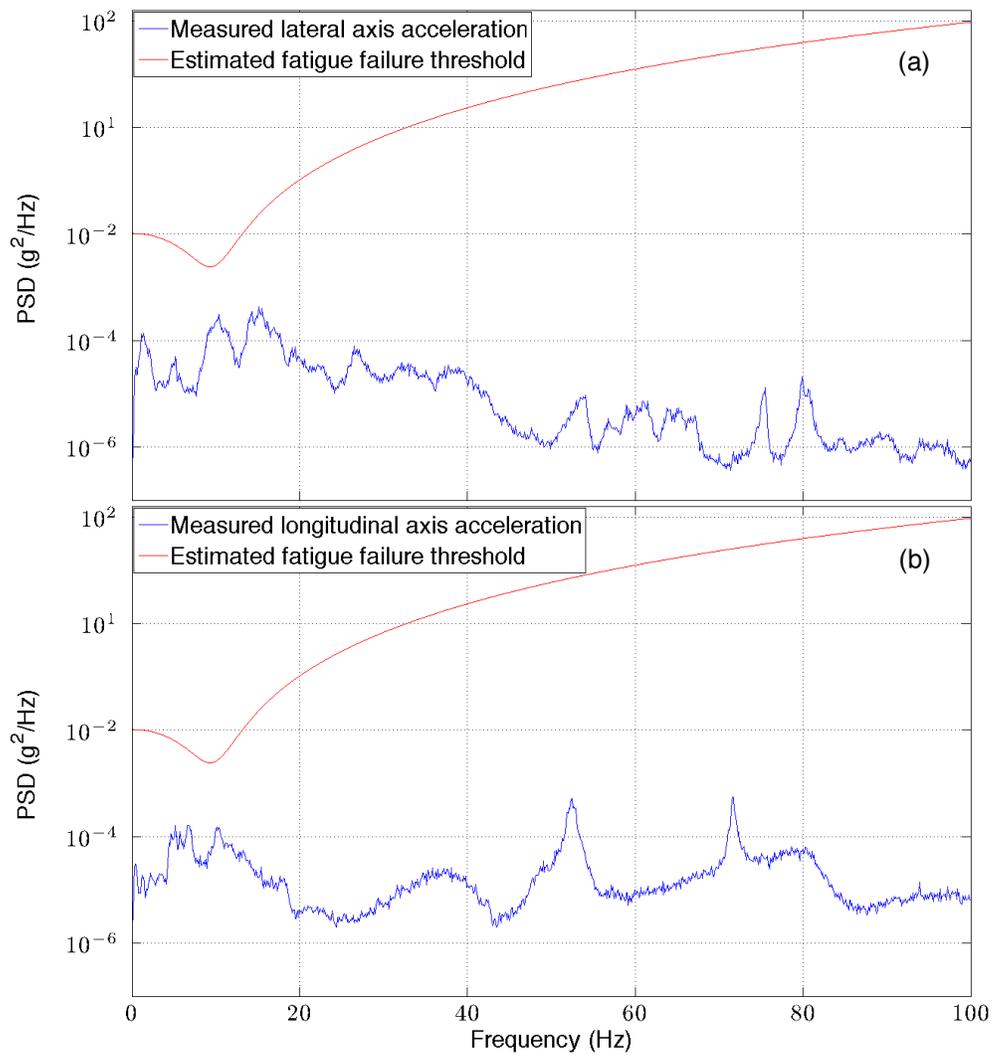

Fig. 7: Acceleration PSD of (a) lateral and (b) longitudinal axis acceleration for typical smooth road conditions. Also shown is the PSD form of the acceleration at the fatigue failure threshold.

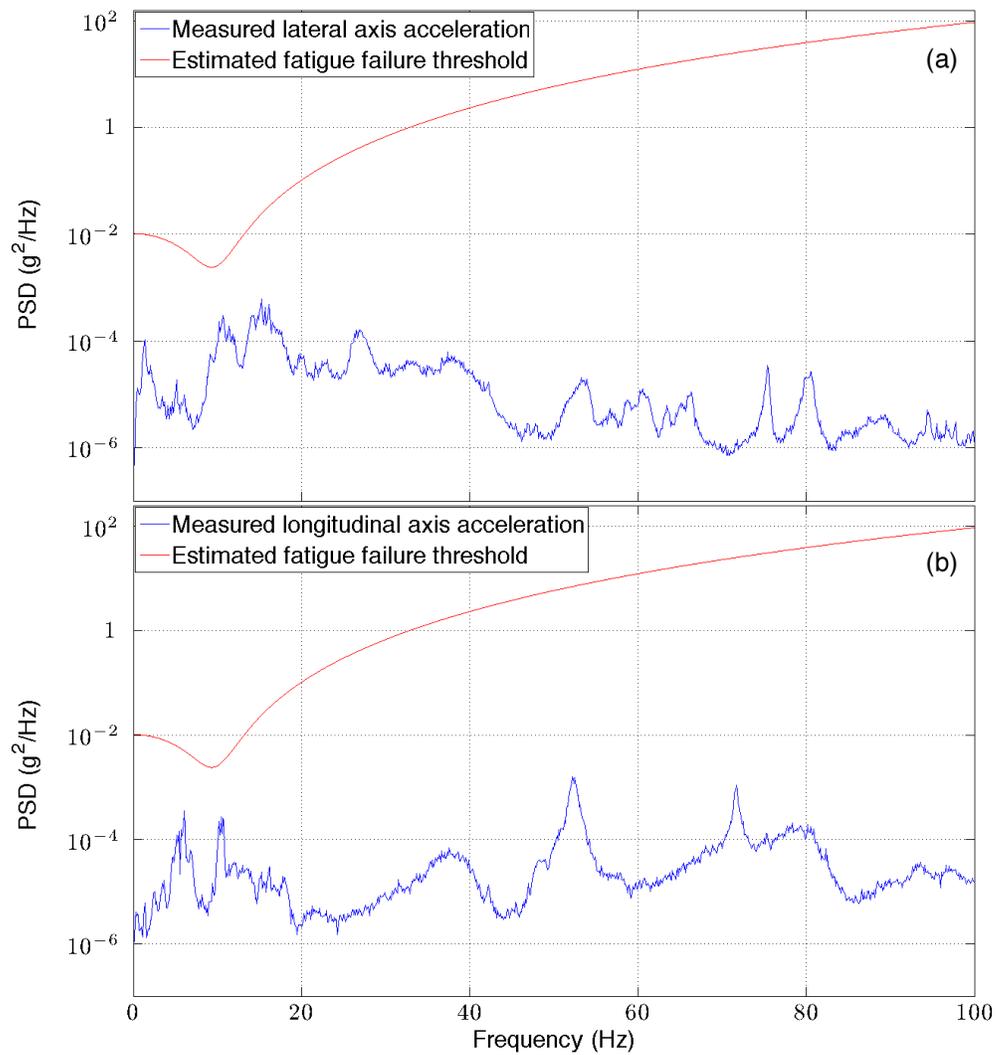

Fig. 8: Acceleration PSD of (a) lateral and (b) longitudinal axis acceleration for typical rough road conditions. Also shown is the PSD form of the acceleration at the fatigue failure threshold.

A comparison of the lateral and longitudinal power spectra for smooth and rough road conditions shows a 20% increase in the average amplitude of oscillations on the rough road sections (Fig. 9). The oscillation amplitude on a rough road (as perceived by the support crew) is greater than that of the smooth road only at frequencies above around 30 Hz.

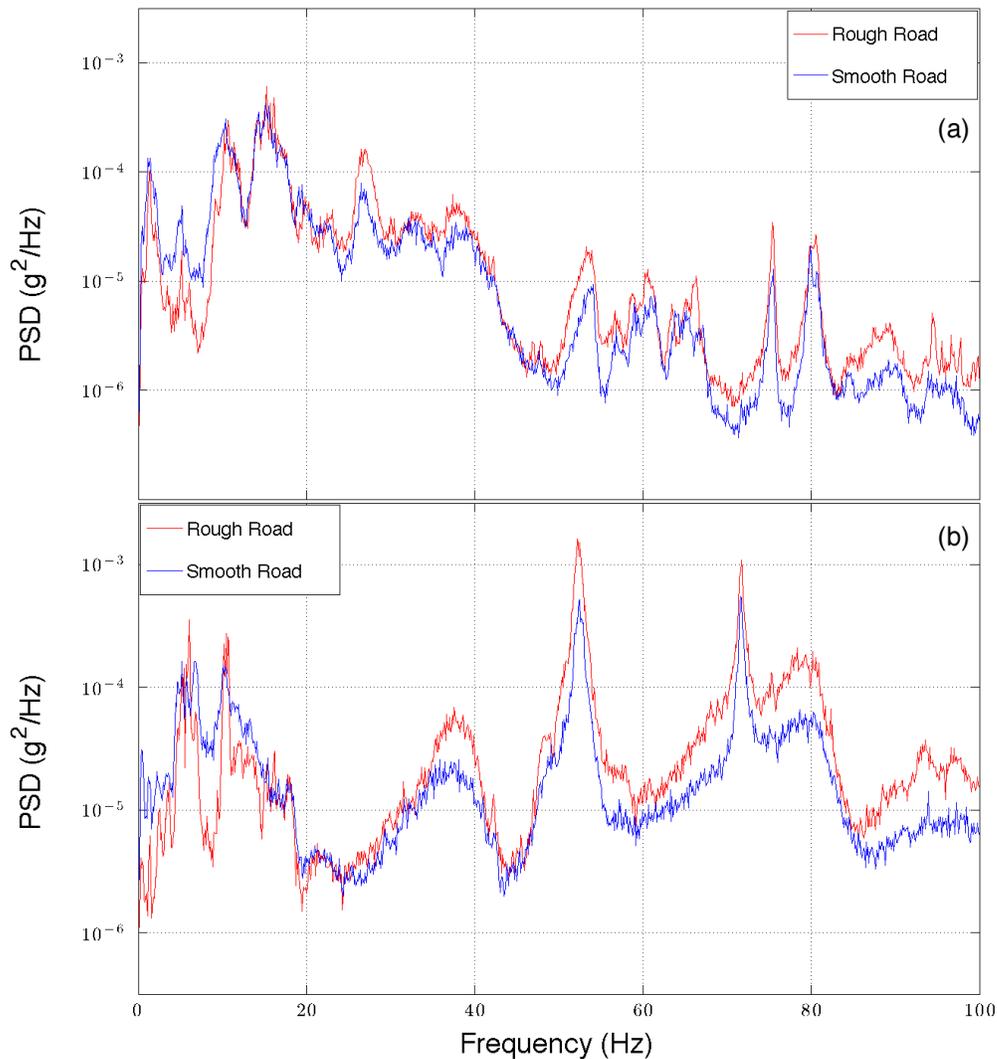

Fig. 9: Comparison of power spectrum for smooth and rough road conditions on (a) lateral and (b) longitudinal axis.

## V. Conclusion

The cryostat arrived safely at the University of Washington after 30 hours of driving time over a 3-day period. A heavy-duty forklift was used to transfer the cryostat from the trailer to the experiment hall where a visual inspection revealed no damage. As a final test, the superconducting magnet coil was cooled with liquid helium and ramped to operating current with no faults, verifying the success of the move.

The test move played a large role in the planning and success of the move. It provided accurate estimates of the magnitude and frequency of the accelerations that would be applied to the cryostat and established logistical guidelines to ensure as few surprises as possible along the way.

The ADMX collaboration gratefully acknowledges support by the state of Washington under budget number 74-0474 SULRMV and the U.S. Department of Energy and Office of High Energy Physics under contract numbers DE-AC52- 07NA27344 (Lawrence Livermore National Laboratory), and DE-FG02-97ER41029 (University of Florida). Additional support was provided by Lawrence Livermore National Laboratory under the LDRD program. This research was supported by the Director, Office of Science, Office of Basic Energy Sciences, Materials Sciences and Engineering Division, of the U.S. Department of Energy under Contract No. DE-AC03-76SF00098 (JC).

| Axis | Max (g) | Avg (g) |
|---|---|---|
| Lateral | 0.580 | 0.031 |
| Longitudinal | 1.510 | 0.057 |
| Vertical | 0.880 | 0.045 |

Table 1 - Test move data statistics for maximum (Max) and average (Avg) accelerations.

| Axis | Max (g) | Avg (g) |
|---|---|---|
| **Overall** | | |
| Lateral | 1.255 | 0.038 |
| Longitudinal | 2.108 | 0.043 |
| **Smooth Road Conditions** | | |
| Lateral | 0.659 | 0.034 |
| Longitudinal | 0.410 | 0.038 |
| **Rough Road Conditions** | | |
| Lateral | 1.018 | 0.035 |
| Longitudinal | 0.411 | 0.050 |

Table 2 - Move data statistics for maximum (Max) and average (Ave) accelerations.